\PassOptionsToPackage{unicode}{hyperref}
\PassOptionsToPackage{hyphens}{url}
\PassOptionsToPackage{dvipsnames,svgnames,x11names}{xcolor}
\documentclass[
  letterpaper,
  DIV=11,
  numbers=noendperiod]{scrartcl}

\usepackage{amsmath,amssymb}
\usepackage{iftex}
\ifPDFTeX
  \usepackage[T1]{fontenc}
  \usepackage[utf8]{inputenc}
  \usepackage{textcomp} 
\fi
\usepackage{lmodern}
\ifPDFTeX\else  
\fi
\IfFileExists{upquote.sty}{\usepackage{upquote}}{}
\IfFileExists{microtype.sty}{
  \usepackage[]{microtype}
  \UseMicrotypeSet[protrusion]{basicmath} 
}{}
\makeatletter
\@ifundefined{KOMAClassName}{
  \IfFileExists{parskip.sty}{%
    \usepackage{parskip}
  }{
    \setlength{\parindent}{0pt}
    \setlength{\parskip}{6pt plus 2pt minus 1pt}}
}{
  \KOMAoptions{parskip=half}}
\makeatother
\usepackage{xcolor}
\makeatletter
\ifx\paragraph\undefined\else
  \let\oldparagraph\paragraph
  \renewcommand{\paragraph}{
    \@ifstar
      \xxxParagraphStar
      \xxxParagraphNoStar
  }
  \newcommand{\xxxParagraphStar}[1]{\oldparagraph*{#1}\mbox{}}
  \newcommand{\xxxParagraphNoStar}[1]{\oldparagraph{#1}\mbox{}}
\fi
\ifx\subparagraph\undefined\else
  \let\oldsubparagraph\subparagraph
  \renewcommand{\subparagraph}{
    \@ifstar
      \xxxSubParagraphStar
      \xxxSubParagraphNoStar
  }
  \newcommand{\xxxSubParagraphStar}[1]{\oldsubparagraph*{#1}\mbox{}}
  \newcommand{\xxxSubParagraphNoStar}[1]{\oldsubparagraph{#1}\mbox{}}
\fi
\makeatother

\providecommand{\tightlist}{%
  \setlength{\itemsep}{0pt}\setlength{\parskip}{0pt}}\usepackage{longtable,booktabs,array}
\usepackage{calc} 
\usepackage{etoolbox}
\makeatletter
\patchcmd\longtable{\par}{\if@noskipsec\mbox{}\fi\par}{}{}
\makeatother
\IfFileExists{footnotehyper.sty}{\usepackage{footnotehyper}}{\usepackage{footnote}}
\makesavenoteenv{longtable}
\usepackage{graphicx}
\makeatletter
\newsavebox\pandoc@box
\newcommand*\pandocbounded[1]{
  \sbox\pandoc@box{#1}%
  \Gscale@div\@tempa{\textheight}{\dimexpr\ht\pandoc@box+\dp\pandoc@box\relax}%
  \Gscale@div\@tempb{\linewidth}{\wd\pandoc@box}%
  \ifdim\@tempb\p@<\@tempa\p@\let\@tempa\@tempb\fi
  \ifdim\@tempa\p@<\p@\scalebox{\@tempa}{\usebox\pandoc@box}%
  \else\usebox{\pandoc@box}%
  \fi%
}
\def\fps@figure{htbp}
\makeatother
\NewDocumentCommand\citeproctext{}{}

\makeatletter
 \let\@cite@ofmt\@firstofone
 \def\@biblabel#1{}
 \def\@cite#1#2{{#1\if@tempswa , #2\fi}}
\makeatother
\newlength{\cslhangindent}
\newlength{\csllabelwidth}
\newenvironment{CSLReferences}[2] 
 {\begin{list}{}{%
  \setlength{\itemindent}{0pt}
  \setlength{\leftmargin}{0pt}
  \setlength{\parsep}{0pt}
  \ifodd #1
   \setlength{\leftmargin}{\cslhangindent}
   \setlength{\itemindent}{-1\cslhangindent}
  \fi
  \setlength{\itemsep}{#2\baselineskip}}}
 {\end{list}}
\usepackage{calc}

\newcommand{\CSLLeftMargin}[1]{\parbox[t]{\csllabelwidth}{\strut#1\strut}}
\newcommand{\CSLRightInline}[1]{\parbox[t]{\linewidth - \csllabelwidth}{\strut#1\strut}}

\KOMAoption{captions}{tableheading}
\makeatletter
\@ifpackageloaded{caption}{}{\usepackage{caption}}
\AtBeginDocument{%
\ifdefined\contentsname
  \renewcommand*\contentsname{Table of contents}
\else
  \newcommand\contentsname{Table of contents}
\fi
\ifdefined\listfigurename
  \renewcommand*\listfigurename{List of Figures}
\else
  \newcommand\listfigurename{List of Figures}
\fi
\ifdefined\listtablename
  \renewcommand*\listtablename{List of Tables}
\else
  \newcommand\listtablename{List of Tables}
\fi
\ifdefined\figurename
  \renewcommand*\figurename{Figure}
\else
  \newcommand\figurename{Figure}
\fi
\ifdefined\tablename
  \renewcommand*\tablename{Table}
\else
  \newcommand\tablename{Table}
\fi
}
\@ifpackageloaded{float}{}{\usepackage{float}}
\floatstyle{ruled}
\@ifundefined{c@chapter}{\newfloat{codelisting}{h}{lop}}{\newfloat{codelisting}{h}{lop}[chapter]}
\floatname{codelisting}{Listing}

\makeatother
\makeatletter
\@ifpackageloaded{caption}{}{\usepackage{caption}}
\@ifpackageloaded{subcaption}{}{\usepackage{subcaption}}
\makeatother

\usepackage{bookmark}

\IfFileExists{xurl.sty}{\usepackage{xurl}}{} 
\hypersetup{
  pdftitle={The Lipid Interactome},
  pdfkeywords={Lipid-protein
interactions, Proteomics, Interactomics, Database, Multifunctionalized
lipids},
  colorlinks=true,
  linkcolor={blue},
  filecolor={Maroon},
  citecolor={Blue},
  urlcolor={Blue},
  pdfcreator={LaTeX via pandoc}}

\title{The Lipid Interactome}
\usepackage{etoolbox}
\makeatletter
\providecommand{\subtitle}[1]{
  \apptocmd{\@title}{\par {\large #1 \par}}{}{}
}
\makeatother
\subtitle{An interactive and open access platform for exploring cellular
lipid-protein interactomes}
\author{Gaelen Guzman \and André Nadler \and Frank Stein \and Jeremy M.
Baskin \and Carsten Schultz \and Fikadu Tafesse}
\date{2025-02-04}

\begin{document}
\maketitle

\begin{center}\rule{0.5\linewidth}{0.5pt}\end{center}

\subsection{Authors}\label{authors}

Gaelen Guzman\textsuperscript{1,}*, André Nadler\textsuperscript{2},
Frank Stein\textsuperscript{3}, Jeremy M. Baskin\textsuperscript{4},
Carsten Schulz\textsuperscript{1}, Fikadu G. Tafesse\textsuperscript{1}

\textsuperscript{1} = Departments of Molecular Immunology and Immunology
and Chemical Physiology and Biochemistry, Oregon Health \& Science
University, Portland, OR, USA; \textsuperscript{2} = Max Plank Institute
of Molecular Cell Biology and Genetics, Dresden Germany;
\textsuperscript{3} = Proteomics Core Facility, European Molecular
Biology Laboratory, Heidelberg, Germany; \textsuperscript{4} =
Department of Chemistry and Chemical Biology and Weill Institute for
Cell and Molecular Biology, Cornell University, Ithaca, NY, USA; * =
Corresponding author:
\href{mailto:gaelen.guzman@lipidinteractome.org}{\nolinkurl{gaelen.guzman@lipidinteractome.org}}

\subsection{Summary}\label{summary}

\subsubsection{Abstract}\label{abstract}

Lipid--protein interactions play essential roles in cellular signaling
and membrane dynamics, yet their systematic characterization has long
been hindered by the inherent biochemical properties of lipids. Recent
advances in functionalized lipid probes -- equipped with
photoactivatable crosslinkers, affinity handles, and photocleavable
protecting groups -- have enabled proteomics-based identification of
lipid interacting proteins with unprecedented specificity and
resolution. Despite the growing number of published lipid interactomes,
there remains no centralized effort to harmonize, compare, or integrate
these datasets.

The Lipid Interactome addresses this gap by providing a structured,
interactive web portal that adheres to FAIR data principles -- ensuring
that lipid interactome studies are Findable, Accessible, Interoperable,
and Reusable. Through standardized data formatting, interactive
visualizations, and direct cross-study comparisons, this resource
enables researchers to systematically explore the protein-binding
partners of diverse bioactive lipids. By consolidating and curating
lipid interactome proteomics data from multiple studies, the Lipid
Interactome database serves as a critical tool for deciphering the
biological functions of lipids in cellular systems.

\subsubsection{Availability}\label{availability}

This site can be viewed at
\href{lipidinteractome.org}{LipidInteractome.org} all data is available
for download. No user information is collected or necessary for data
navigation, interaction, or download.

\subsubsection{Data Submission \&
Contact}\label{data-submission-contact}

If you would like to submit a relevant dataset, see our
\href{https://lipidinteractome.org/contactus/datasubmission}{Data
Submission Page} for details, contact info, and instructions. For site
feedback and additional questions, please email the site administrators
at
\href{mailto:Contact.Us@lipidinteractome.org}{\nolinkurl{Contact.Us@lipidinteractome.org}}

\subsubsection{Keywords}\label{keywords}

Lipid-protein interactions, proteomics, interactomics, database,
multifunctionalized lipids, photoaffinity labeling

\subsection{Introduction}\label{introduction}

Historically, lipids have been considered merely the passive structural
components of cellular membranes. However, recent decades have revealed
that this diverse category of biomolecule plays essential and active
roles in the cell, ranging from modulating membrane dynamics,
determining organelle identity, serving as catabolites for energy
storage and metabolism, and bioactive signaling mediators for organellar
and intracellular signaling\textsuperscript{1--6}. Interactions with
proteins are essential to many of these crucial lipid functions -- and
yet, lipid--protein interactions are extremely challenging to study: The
small size and biophysical properties of lipids make conventional
proteomics approaches unsuitable for mapping lipid
interactors\textsuperscript{7}. Unlike proteins, which can be readily
tagged or immobilized for affinity-based interaction studies, lipids
require specialized chemical modifications to enable selective and
controlled interaction profiling\textsuperscript{8}.

Recent advances in synthetic, multifunctionalized lipid analogs have
revolutionized the field, enabling the systematic interrogation of lipid
interactomes\textsuperscript{8}. These probes incorporate strategically
designed functional groups that enable covalent crosslinking, affinity
enrichment, and controlled activation. Most commonly, these include: (1)
a diazirine group for photoactivatable crosslinking upon exposure to 355
nm ultraviolet (UV) light, stabilizing non-covalent and often
modest-affinity lipid--protein interactions; (2) a terminal alkyne,
permitting click chemistry-based enrichment; and, in some cases, (3) a
coumarin photocage, which shields the lipid probe from premature
enzymatic modification until it is selectively cleaved by 405 nm light
(a wavelength that does not induce diazirine
photolysis)\textsuperscript{7--10}. Together, these modifications enable
direct proteomic identification of lipid interactors, distinguishing
true binding partners from transient or non-specific
associations\textsuperscript{7,10,11}.

Numerous studies have already employed these lipid probes to elucidate
the interactomes of key bioactive lipids, including diacylglycerol,
sphingosine, sphinganine, phosphatidic acid, and
phosphatidylethanolamine, N-acylphosphatidylethanolamine, and
phosphatidyl alcohol\textsuperscript{9,11--16}. These initial studies
have followed a comparative proteomics workflow, wherein crosslinked
(``+UV'') and non-crosslinked (``-UV'') samples are analyzed side by
side to minimize artifacts and confirm
specificity\textsuperscript{7,10}. However, despite the growing number
of lipid interactome datasets, no centralized resource has existed to
compile, compare, and standardize these findings.

To address this gap, we present the
\href{https://www.lipidinteractome.org}{Lipid Interactome} -- a curated
repository designed to facilitate the comparative analysis of
lipid--protein interactions from multiple studies. This resource
consolidates proteomics data from multifunctional lipid probes, enabling
researchers to efficiently explore and contextualize lipid-binding
proteins across different lipid species and cellular models. Currently,
the repository hosts interactomes derived from trifunctionalized
phosphatidic acid, phosphatidylethanolamine, phosphatidylinositol
\emph{tris}phosphate, N-acylphosphatidylethanolamine, phosphatidyl
alcohol, sphingosine, and sphinganine in HeLa and Huh7
cells\textsuperscript{11,13--16}. Additionally, it includes control
datasets from two fatty acid probes to distinguish specific from
non-specific lipid-binding proteins\textsuperscript{12}. We additionally
intend to expand this resource to include more complex experimental
designs as they are published (e.g.~Disease versus Healthy Control or
Infection versus Mock).

By providing a centralized platform for lipid interactome data, the
Lipid Interactome significantly lowers the barriers to entry for
researchers investigating lipid signaling pathways, enabling hypothesis
generation through comparative analysis of multiple studies. Overall,
this repository represents an important step toward systematizing
lipid--protein interaction studies, ensuring that bioactive lipids are
afforded the same level of proteomic scrutiny as their protein
counterparts.

\subsection{Description}\label{description}

\subsubsection{Overview}\label{overview}

The~Lipid Interactome~is an interactive discovery platform built
using~Quarto, designed to centralize and enhance the study of
lipid--protein interactions through curated datasets and dynamic
visualization tools. Each page integrates~R-generated figures, embedding
scripts directly into the source material to ensure reproducibility and
transparency. The repository currently hosts data from~seven published
studies, with an expandable framework that supports the incorporation of
future datasets as the field progresses.

At its core, the~Lipid Interactome~enables the~direct, qualitative
comparison~of lipid interactomes between studies, providing researchers
with invaluable insights into the diverse roles that bioactive lipids
play in cellular processes. By leveraging standardized data formatting
and interactive visualization -- including study-specific analysis
pages, lipid probe aggregation, and cross-study comparisons via
a~Shiny~application -- the platform facilitates intuitive exploration of
protein--lipid interactions. As a~living repository, we actively
encourage researchers to contribute their findings through
our~\href{https://lipidinteractome.org/contactus/datasubmission}{Data~Submission}~page,
ensuring that this resource continues to evolve alongside advancements
in lipidomics research.

\subsubsection{FAIR data principles}\label{fair-data-principles}

The~Lipid Interactome~database was developed with the FAIR (Findable,
Accessible, Interoperable, and Reusable) data principles in mind,
ensuring that lipid--protein interaction data can be systematically
explored, compared, and reused by the research community. Given the
historical inconsistency in proteomics data curation, we have taken
deliberate steps to standardize, annotate, and document our datasets for
maximum utility.

\paragraph{Findable}\label{findable}

To address the fragmented nature of lipid interactome studies, we have
established a centralized platform where researchers can interact with
the results of individual studies or explore aggregated findings across
multiple studies on a given lipid. The repository is structured to
facilitate intuitive navigation between experimental datasets, offering
clear entry points for study-specific and lipid-specific analyses.
Because this resource is intended as a living repository, we actively
invite researchers to contribute their findings following the submission
guidelines detailed on
our~\href{https://lipidinteractome.org/contactus/datasubmission}{Data
Submission}~page.

\paragraph{Accessible}\label{accessible}

All datasets within~the Lipid Interactome~are available for download in
standardized~\texttt{.csv}~format, ensuring broad accessibility across
different computational environments. To further enhance usability, we
provide
a~\href{https://www.lipidinteractome.org/background/multifunctionallipidprobesoverview}{detailed~overview~of
data~generation~and~analysis}, including explicit descriptions of
dataset structure, column definitions, and the inherent limitations of
multifunctional lipid probes. Only peer-reviewed proteomics datasets are
included in this repository to maintain high data quality and
reliability.

\paragraph{Interoperable}\label{interoperable}

Given the diversity of methodologies employed across lipid interactome
studies, we have taken care to harmonize data formats and metadata
annotations. While many studies share common analytical pipelines,
others employ distinct mass spectrometry workflows. To ensure
consistency, we have structured the data to allow parallel comparisons
and
provided~\href{https://www.lipidinteractome.org/background/proteomicsusingmultifunctionalprobes}{documentation~on~methodological~differences}.
This standardization effort facilitates direct cross-study comparisons
without requiring extensive reformatting by the end user. Each dataset
is accompanied by metadata which describe the sample handling and data
analysis techniques -- with clarifying text describing how
``significance'' is determined for each analysis.

\paragraph{Reusable}\label{reusable}

To maximize the utility of these datasets, we provide extensive
documentation detailing data acquisition, analysis workflows, and
potential limitations. Wherever possible, technical jargon has been
minimized to ensure accessibility to a broad audience, from lipid
biologists to computational scientists. Additionally, we reference key
publications that delve deeper into the synthesis of lipid probes and
the specifics of each proteomics workflow, allowing researchers to
further contextualize and validate their findings.

By adhering to these FAIR principles,~the Lipid Interactome~database
aims to serve as a robust and enduring resource for the lipid biology
community, reducing redundancy, improving data transparency, and
accelerating the study of lipid-mediated signaling.

\subsubsection{Data Visualization and Download
Modules}\label{data-visualization-and-download-modules}

To facilitate efficient exploration and analysis of lipid--protein
interaction data, the~Lipid Interactome~repository incorporates a suite
of interactive visualization and data access tools. These modules are
designed to enhance user engagement, provide clear interpretability of
experimental findings, and streamline data retrieval for further
computational analysis -- overall, the site is designed to ensure that
the~Lipid Interactome~is not merely a static repository, but rather a
dynamic and evolving resource for lipid biologists and bioinformaticians
alike.

\paragraph{Study-Specific Data
Exploration}\label{study-specific-data-exploration}

Each publication contributing to the repository is provided with a
dedicated results page, allowing users to examine study findings in
detail. These pages include interactive~Volcano,~Ranked-Order, and~MA
plots, which enable visualization of differential protein enrichment
across experimental conditions. The plots are rendered using
the~Plotly~and~htmlwidgetsR packages, ensuring a responsive, client-side
experience that supports zooming, hovering, image export, and data
filtering\textsuperscript{17,18}. For transparency and reproducibility,
all datasets associated with a given study are available for direct
download.

\paragraph{Lipid Probe-Specific Pages}\label{lipid-probe-specific-pages}

In addition to the pages for each independent publication, datasets are
further organized by lipid probe, providing a consolidated view of
protein interactors detected across multiple experiments and independent
studies. Each lipid-specific page includes study metadata and cell type
-- facilitating cross-study comparisons. Future iterations of the
repository will extend this functionality by enabling direct comparisons
of interactomes obtained from different cell lines treated with the same
lipid probe, further refining the contextual understanding of
lipid--protein interactions.

\paragraph{Comparative Analysis via Shiny
Application}\label{comparative-analysis-via-shiny-application}

Where study design permits, data has been integrated into an
interactive~Shiny~application, allowing users to perform direct (though
qualitative)~probe-to-probe
comparisons~(\href{https://www.lipidinteractome.org/lipidprobe/enrichedhitscomparison}{Probe~vs~Probe~Comparisons}).
This tool enables researchers to explore the similarity of lipid
interactomes across experimental conditions via~linear regression
analysis, highlighting proteins that are significantly enriched in one
dataset, both datasets, or neither. By providing an intuitive framework
for interactome comparison, this feature aids in the identification of
promising targets for further investigation.

Because each publication utilizes distinct sample handling and data
analysis techniques, we emphasize that these comparisons are qualitative
-- we hope to discourage quantitative statements regarding relative
enrichment between lipid probes (e.g.~``Protein X is enriched 3x more to
Lipid A than to Lipid B''). However, great utility lies in assessing
co-enrichment (or orthogonal enrichment) of proteins to disparate lipid
probes without quantitative comparison.

\subsubsection{Data format}\label{data-format}

While there are numerous ways to quantify the mass spectrometry (MS)
data, the majority of studies included in our repository have utilized
the same (or highly similar) quantitative MS pipeline using
data-dependent acquisition techniques, such as Tandem Mass
Tagging\textsuperscript{19}. These datasets include those reported by
Farley et al.~(2024), Farley et al.~(2024), and Thomas et
al.~(2025)\textsuperscript{11--13}. Some studies have also featured
quantitative MS analysis using stable isotope labeling of amino acids in
cell culture (SILAC), which also enables extraction of key data
described below\textsuperscript{14,15}. Correspondingly, the ideal data
structure for incorporation into our repository will include the
following:

\begin{longtable}[]{@{}
  >{\raggedright\arraybackslash}p{(\linewidth - 10\tabcolsep) * \real{0.1486}}
  >{\centering\arraybackslash}p{(\linewidth - 10\tabcolsep) * \real{0.1081}}
  >{\centering\arraybackslash}p{(\linewidth - 10\tabcolsep) * \real{0.3514}}
  >{\centering\arraybackslash}p{(\linewidth - 10\tabcolsep) * \real{0.1216}}
  >{\centering\arraybackslash}p{(\linewidth - 10\tabcolsep) * \real{0.1622}}
  >{\centering\arraybackslash}p{(\linewidth - 10\tabcolsep) * \real{0.1081}}@{}}
\caption{\textbf{Example data table for smooth incorporation into Lipid
Interactome.} \textbf{\emph{Column descriptions:}} \emph{Gene name}: The
identifier for genes identified in the dataset (or Accession Number).
\emph{Uniprot accession}: Stable identifiers of UniProtKB entries;
integrates species and isoform data on the identified protein.
\emph{Enrichment value}: Fold change of Experimental versus Control
fold-change, typically log2-transformed -- ideally the product of TMT or
other quantitative mass spectrometry methods. \emph{p value}: the
results of an ANOVA or LIMMA analysis, corrected for multiple hypothesis
testing. \emph{AveExpr}: An averaged value of the m/z ion intensity of
the Experimental condition and Control condition used for the enrichment
value (used to prepare MA plots for quality control). \emph{FDR}:
False-discovery rate of protein
identification.}\label{tbl-1}\tabularnewline
\toprule\noalign{}
\begin{minipage}[b]{\linewidth}\raggedright
Gene name
\end{minipage} & \begin{minipage}[b]{\linewidth}\centering
\href{https://www.uniprot.org/help/accession_numbers}{Uniprot accession}
\end{minipage} & \begin{minipage}[b]{\linewidth}\centering
logFC
\end{minipage} & \begin{minipage}[b]{\linewidth}\centering
p value
\end{minipage} & \begin{minipage}[b]{\linewidth}\centering
AveExpr
\end{minipage} & \begin{minipage}[b]{\linewidth}\centering
FDR
\end{minipage} \\
\midrule\noalign{}
\endfirsthead
\toprule\noalign{}
\begin{minipage}[b]{\linewidth}\raggedright
Gene name
\end{minipage} & \begin{minipage}[b]{\linewidth}\centering
\href{https://www.uniprot.org/help/accession_numbers}{Uniprot accession}
\end{minipage} & \begin{minipage}[b]{\linewidth}\centering
logFC
\end{minipage} & \begin{minipage}[b]{\linewidth}\centering
p value
\end{minipage} & \begin{minipage}[b]{\linewidth}\centering
AveExpr
\end{minipage} & \begin{minipage}[b]{\linewidth}\centering
FDR
\end{minipage} \\
\midrule\noalign{}
\endhead
\bottomrule\noalign{}
\endlastfoot
GeneA & P80723 & 1.2 & 0.001 & 1.0x10\textsuperscript{18} & 0.0001 \\
GeneB & P32004 & 0.1 & 0.6 & 1.0x10\textsuperscript{5} & 0.005 \\
\ldots{} & \ldots{} & \ldots{} & \ldots{} & \ldots{} & \ldots{} \\
\end{longtable}

\textbf{We wish to emphasize that we will collaborate with data
providers to prepare visualizations which best suit their datasets --
the example dataset above merely depicts an ideal case which allows for
easy incorporation and cross-study comparison.}

Additionally, we are working on providing visualization functionality
for studies which utilized Peptide Spectral Matching (PSM) to report
enrichment toward lipid probes. Similarly, data-independent acquisitio
(DIA) is a quickly emerging alternative to data-dependent acquisition
(DDA; e.g.~TMT)\textsuperscript{20,21}; as with PSM datasets, we are
working to incorporate DIA data formats.

\subsection{Discussion}\label{discussion}

The advent of functionalized lipid probes has fundamentally transformed
lipid biology by enabling the proteomic identification of lipid-binding
proteins with unprecedented specificity. However, as the number of
published lipid interactome studies grows, the lack of a centralized
framework for comparing and integrating these datasets presents a
significant challenge for deciphering biological functions of individual
lipids and lipid-binding proteins holistically and identifying trends
across datasets. The Lipid Interactome was developed to address this
need by providing researchers with a structured platform to explore,
cross-reference, and build upon previous findings. By aggregating and
standardizing lipid--protein interaction data, this resource facilitates
the rapid identification of candidate lipid-binding proteins and
supports orthogonal validation efforts.

Designed as a continuously expanding repository, the Lipid Interactome
will evolve alongside advances in lipid probe technologies and
interactomics. Future iterations will incorporate validation studies for
each lipid probe, offering researchers a dedicated space to compare
findings across experimental contexts. This functionality will not only
reduce redundancy in experimental design but also strengthen the
reproducibility and biological interpretation of lipid interactome
research. To support this vision, we invite researchers to contribute
their datasets via our
\href{https://lipidinteractome.org/contactus/datasubmission}{Data
Submission page}, fostering a collaborative and transparent approach to
lipid biology.

\subsection{Abbreviations}\label{abbreviations}

\begin{description}
\tightlist
\item[UV]
\emph{U}ltra\emph{V}iolet
\item[MS]
\emph{M}ass \emph{S}pectrometry
\item[LIMMA]
\emph{Li}near \emph{M}odels for \emph{M}icro\emph{a}rray and RNA-Seq
Data
\item[ANOVA]
\emph{An}alysis \emph{o}f \emph{Va}riance
\item[PSM]
\emph{P}eptide \emph{S}pectral \emph{M}atching
\item[DIA]
\emph{D}ata-\emph{I}ndedpendent \emph{A}cquisition
\item[DDA]
\emph{D}ata-\emph{D}edpendent \emph{A}cquisition
\end{description}

\subsection{Contributions}\label{contributions}

GG designed, implemented, and wrote the majority of the LipidInteractome
repository and manuscript. FS wrote the data analysis section of the
repository, provided significant intellectual contribution to the
repository design, and edited the manuscript. AN provided substantial
editing of the repository and manuscript. JMB edited the manuscript and
contributed data to the repository. CS and FT provided significant
intellectual contribution and text editing of the repository and
manuscript and oversaw the site development.

\subsection{Funding}\label{funding}

This work was supported by funding from the National Institutes of
Health (NIAID R01 AI141549-02, to FGT, CS; NIGMS R01 GM127631 to C.S.,
and NIGMS R01 GM151682 to JMB).

\subsection{Conflicts of Interest}\label{conflicts-of-interest}

The authors have no conflicts of interest to declare.

\subsection{Citations}\label{citations}

\phantomsection\label{refs}

\end{document}